\documentclass[twocolumn,showpacs,reprint,prb,aps,superscriptaddress]{revtex4-2}
\usepackage{amsmath}
\usepackage{amssymb}
\usepackage[retain-unity-mantissa=false]{siunitx}
\usepackage{graphicx}
\usepackage{placeins}
\usepackage[colorlinks=true, linkcolor=blue, citecolor=blue, filecolor=blue, urlcolor=blue,pdfproducer={.},pdfcreator={.}]{hyperref}
\usepackage{cleveref}
\usepackage{blindtext}
\usepackage{tikz}
\usepackage{verbatim}
\usepackage{ulem}
\usepackage[commandnameprefix=always, commentmarkup=uwave]{changes}
\usepackage{physics}
\usepackage{colortbl}

\newcommand{\els}{_\text{el*}}
\newcommand{\el}{_\text{el}}
\newcommand{\ep}{_\text{ep}}
\newcommand{\ph}{_\text{ph}}
\newcommand{\elel}{_\text{el*-el}}
\newcommand{\elsel}{_\text{el*-el}} %same but clearer definition, the order matters!
\newcommand{\elels}{_\text{el-el*}}
\newcommand{\elph}{_\text{el*-ph}}
\newcommand{\elsph}{_\text{el*-ph}} % better naming
\newcommand{\phels}{_\text{ph-el*}}

\newcommand{\de}{{\rm d}E}

\newcommand{\ii}[1]{_\text{#1}}
\newcommand{\rel}{^\text{rlx}}

\begin{document}
\title{Capturing non-equilibrium electron dynamics in metals accurately and efficiently}
	   \author{M. \surname{Uehlein}}
	\affiliation{Department of Physics and Research Center OPTIMAS, RPTU Kaiserslautern-Landau, 67663 Kaiserslautern, Germany}
 
	   \author{H. T. \surname{Snowden}}
	\affiliation{Department of Chemistry, University of Warwick, CV4 7AL Coventry, United Kingdom}

        \author{C. \surname{Seibel}}	
	\affiliation{Department of Physics and Research Center OPTIMAS, RPTU Kaiserslautern-Landau, 67663 Kaiserslautern, Germany}
 
        \author{T. \surname{Held}}	
	\affiliation{Department of Physics and Research Center OPTIMAS, RPTU Kaiserslautern-Landau, 67663 Kaiserslautern, Germany}
 
        \author{S. T. \surname{Weber}}	
	\affiliation{Department of Physics and Research Center OPTIMAS, RPTU Kaiserslautern-Landau, 67663 Kaiserslautern, Germany}
 
        \author{R. J. \surname{Maurer}}
     \email{r.maurer@warwick.ac.uk}
     \affiliation{Department of Chemistry, University of Warwick, CV4 7AL Coventry, United Kingdom}
     \affiliation{Department of Physics, University of Warwick, CV4 7AL Coventry, United Kingdom}
 
	\author{B. \surname{Rethfeld}}
 \email{rethfeld@rptu.de}
	\affiliation{Department of Physics and Research Center OPTIMAS, RPTU Kaiserslautern-Landau, 67663 Kaiserslautern, Germany}

	\begin{abstract}
        The simulation of non-equilibrium electron distributions is essential for capturing light-metal interactions and therefore the study of photoabsorption, photocatalysis, laser ablation, and many other phenomena. 
        Current methodologies, such as the Boltzmann equation using full collision integrals, describe non-equilibrium electron dynamics in great detail but at often prohibitive computational expense. 
        In contrast, the simplification via a relaxation time approach can hinder the description of important features or, even worse, lead to nonphysical behavior due to the lack of particle and energy conservation. 
        We propose a model that bridges the gap between the Boltzmann equation and two-temperature models to trace non-equilibrium distributions efficiently. 
        This Athermal Electron Model (AthEM) separately captures 
        the dynamics of thermal and athermal electrons and describes the energy and particle flow between two electronic systems and phonons. 
        We show that the results align well with the results of Boltzmann equation and data from photoemission experiments. 
        The AthEM enables the rapid generation of qualitatively accurate non-equilibrium electron distributions and provides a good starting point for further extensions.  
	\end{abstract}

	\date{\today}

	\maketitle

\section{Introduction}

Exposing metals and semiconductors to light excites electron-hole pairs, generating so-called 'hot carriers', and drives the energy distribution of electrons out of equilibrium. 
Over a timescale of femto- to picoseconds, the material thermalizes and equilibrates through non-equilibrium dynamics that involve electron-electron scattering and electron-phonon scattering. 
Due to rapid electron-electron scattering, electrons relax into a quasi-thermal Fermi-Dirac distribution, which further equilibrates with the phonons in the material~\cite{Mueller2013PRB, brongersma2015plasmon, Rethfeld2002, Pietanza2004}.
The timescale of the electron thermalization is often much faster than the timescale of electron-phonon equilibration, which gave rise to the popular two-temperature model (TTM) \cite{Kaganov1957,Anisimov1974,Rethfeld2017}, where electrons and phonons retain thermal distributions but evolve at different temperatures. However, often, hot/athermal electrons cannot be neglected, and their short-lived presence will significantly affect the measurable spectroscopic response of materials~\cite{lloyd20212021}. Athermal electrons may further induce non-thermal transport or coupled electron-nuclear dynamics, as in photocatalysis or laser-driven ablation~\cite{xiao2020laser,tan2019heterogeneous,maiuri2019ultrafast}. 
Light-driven dynamics depend on many materials' properties and for nanostructured materials and surfaces, many open questions remain on the role of non-thermal carriers in transport, surface dynamics, and electron-hole recombination~\cite{weight2023theory, sivan2024ballistic,stefancu2024electronic}. 
This motivates the continued development of accurate and efficient computational models to simulate how non-equilibrium electron distributions depend on experimental conditions and materials properties.

Beyond the TTM, a range of computational models and methods have been proposed to describe the non-equilibrium dynamics of electrons in materials. \textit{Ab initio} methods such as time-dependent density functional theory (DFT) have been extensively used to study electronic and coupled electron-nuclear dynamics far from equilibrium~\cite{lian2018photoexcitation,herring2023recent,della2022orbital,chu2020long,vanzan2023energy,senanayake2019real,asadi2020td,rossi2017kohn,perera2020plasmonic, Sharma2011, Krieger2015, roman2019}. However, these methods often face limitations in scalability and the ability to capture the relaxation processes that dominate after the laser pulse~\cite{Vanzan2024}. In particular, the commonly employed adiabatic approximation in time-dependent DFT insufficiently captures electron-electron scattering~\cite{lacombe2018electron}. 

One state-of-the-art simulation of laser-excited electron dynamics involves the propagation of the Boltzmann equation~\cite{Sun1994, Pietanza2007, Knorren2000, Ono2018, Caruso2022, DelFatti2000,Krauss2009, Essert2011, Mueller2013PRL,Kaiser2000,Brouwer2017,Rethfeld2002, Mueller2013PRB, Seibel2023, rogier1994direct, Dubi2019, Sun1994, yadav_photocarrier_2019, bringuier_boltzmann_2019,bardeen1958conduction, guenault1964scattering, kabanov2008electron, wilson2020}.
The Boltzmann equation is a master equation for the time evolution of the electronic distribution, which can include dependence on spin and band indices. The rate of change of electron occupation is determined by optical excitation and various scattering effects. The Boltzmann equation with full electron collision integrals can accurately capture thermalization due to electron scattering processes~\cite{Rethfeld2002, Mueller2013PRB, Seibel2023}. Unfortunately, the evaluation of full electron-electron scattering integrals is computationally unfeasible at scale, particularly when spatial transport is considered. This limits its applicability for heterogeneous systems or ultrafast structural dynamics.

Previously, there have been several attempts to simplify the Boltzmann model or extend phenomenological models, such as the TTM.  For example, Boltzmann equation simulations often impose the relaxation time approximation (RTA) for the description of electron-electron scattering or electron-phonon scattering terms~\cite{DelFatti2000, Dubi2019, Sun1994, yadav_photocarrier_2019, bringuier_boltzmann_2019,bardeen1958conduction,guenault1964scattering, kabanov2008electron}.
The RTA replaces scattering integrals with rates that are proportional to relaxation times and differences in distributions~\cite{guenault1964scattering}.
Relaxation times can be assumed to be constant or energy-dependent, with the energy dependence often calculated based on Fermi liquid theory~ \cite{kabanov2008electron,Dubi2019}.
On the other hand, various extensions to incorporate non-equilibrium effects into the TTM have been proposed~\cite{Carpene2006, Tsibidis2018, Uehlein2022, Sun1994, Maldonado2017}. 
Carpene \textit{et al.} introduced an extended two-temperature Model (eTTM)~\cite{Carpene2006}, which was further improved by Tsibidis~\cite{Tsibidis2018} and Uehlein \textit{et~al.}~\cite{Uehlein2022}. 
This model improves upon the TTM by splitting the dynamics of the electronic system into the evolution of thermal electrons and a correction to capture the influence of athermal electrons. 
The model has been shown to give a reliable description of electron dynamics for materials such as bulk aluminum. 
The model assumes a constant density of states and single photon absorption, which limits its applicability to the most simple materials. 

Here, we present a model describing non-equilibrium electron dynamics, which we call Athermal Electron Model (AthEM). 
AthEM explicitly propagates a non-equilibrium electron energy distribution coupled to thermal baths of electrons and phonons (see Figure \ref{fig:panda}). 
The model can be directly motivated based on the Boltzmann equation by separating the electron distribution into a thermal and an athermal contribution. 
The method provides two concrete advantages over previous works: (A) AthEM captures electron-hole pair generation by an external source, such as a laser, via Fermi's golden rule and considers a realistic electronic density of states (DOS). 
(B)  AthEM strictly maintains energy and particle conservation by construction. 
In contrast, previous eTTM variants and RTA applied to Boltzmann equation with energy-dependent relaxation times violate particle conservation.
The AthEM model enables the use of energy-dependent relaxation times instead of full electron-electron collision integrals. This allows AthEM to capture electron-electron scattering accurately while also achieving high computational efficiency.

On the example of gold, we show that AthEM achieves an accurate description of non-equilibrium electron distributions and time-dependent spectral densities. The AthEM model forms thus an excellent starting point for future extensions to study hot electron dynamics, e.g.,  ballistic transport effects.

\section{Results}

The Boltzmann equation in spatially homogeneous materials and without any gradients or external forces reduces to the temporal derivative of the energy distribution of the considered ensemble of particles. 
Different collision processes can affect this distribution. 
If several ensembles of particles are considered, e.g. electrons and phonons, the temporal derivatives of their distributions are coupled through corresponding collision terms. 
In some cases, the details of the distributions are not relevant and the time evolution of electron and phonon distributions can be reduced to the time evolution of effective temperatures for electrons in a Fermi-Dirac distribution and phonons in a Bose-Einstein distribution, which represents the well-established two-temperature model (TTM)~\cite{Anisimov1974}. 
For details regarding the transition from the Boltzmann equation to the TTM, see Supplementary Information (SI), section S1. 

Several implementations of full Boltzmann collision integrals 
have been introduced to describe non-equilibrium electron 
dynamics in materials like noble metals, ferromagnets and dielectrics~
\cite{Rethfeld2002,Caruso2022, Ono2018, DelFatti2000, Knorren2000, Krauss2009, Essert2011, Kaiser2000, Mueller2013PRB, Mueller2013PRL, Brouwer2017}. Among these, several descriptions have considered a separation of electrons into two subsystems~\cite{Mueller2013PRL,Brouwer2017}. 
More details on the general approach to this separation are given in the SI, section~S2.

%%%%%%%%%%%%%%%%%%%%%%%%%%%%%%%%%%%%%%%%%%%%%%%%%%%%%
\subsection{Athermal electron model}

Based on the Boltzmann model for two electronic subsystems and one phonon system,
we develop the Athermal Electron Model (AthEM).
Here, one electronic subsystem is considered to consist of the {\it athermal} electron-hole pairs initially excited by the laser pulse. In contrast, the other electronic subsystem contains only those electrons, which follow a Fermi-Dirac distribution. The phonon system is assumed to follow an equilibrium distribution, here a Bose-Einstein distribution, at all times. 

Athermal excitations decay into the thermal electronic subsystem, increasing its temperature $T\el$ above the temperature of the phonons, $T\ph$, in the process. Thermal electrons and phonons exchange energy
as in the TTM~\cite{Anisimov1974},
and also the athermal electronic subsystem 
transfers energy into the phononic subsystem by electron-phonon collisions.

\begin{figure}[t]
    \centering
    \includegraphics[width=.4\textwidth]{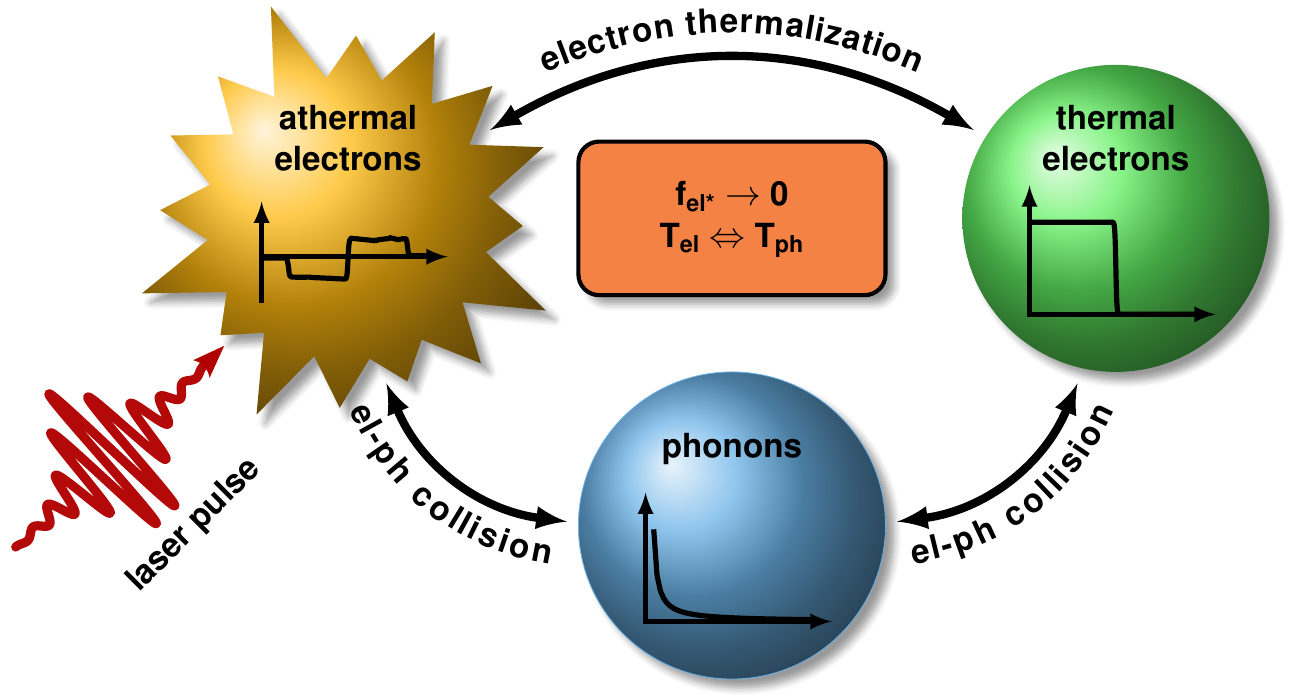}
    \caption{\textbf{Schematic illustration of the three subsystems of the Athermal Electron Model (AthEM).} As in the TTM~\cite{Anisimov1974}, the thermal electrons and phonons are considered as separate subsystems with different temperatures. Due to collisions, they relax to a thermal equilibrium. An optical laser pulse excites the athermal electron subsystem, which thermalizes by collisions with thermal electrons and phonons. The athermal electrons are described by a third subsystem. Figure adapted from~\cite{Uehlein2022}.
    }
    \label{fig:panda}
\end{figure}
Figure~\ref{fig:panda} sketches the considered subsystems and their interactions. Equilibrium is reached when the athermal electron distribution, $ f\els$, where negative values describes holes and positive values describes electrons, has decayed and the thermal electrons and phonons share the same temperature, as indicated with the orange label in the center of the figure.
Similar descriptions have been developed previously, mainly the extended two-temperature model (eTTM)~\cite{Carpene2006,Tsibidis2018,Uehlein2022}. We will refer to the latest implementation of the eTTM by Uehlein {\it et al.}~\cite{Uehlein2022} for comparison below. The main advance between eTTM and AthEM is that the latter improves the description of excitation via Fermi's golden rule and ensures particle conservation (see also SI, section S3).
Moreover, AthEM considers explicitly the dependence of the excitation probability on the DOS. 

The dynamics of the athermal electronic subsystem are given in terms of the temporal changes of its energy distribution, 
\begin{align}
    \pdv{f\els}{t} &= \left.\pdv{f\els}{t}\right|_\text{laser}\!+ \left.\pdv{f\els}{t}\right|\elsel + \left.\pdv{f\els}{t}\right|\elsph\,,\label{eq:f_els}
\end{align}
describing the laser excitation, the changes due to interaction with the thermal electronic subsystem (index 'el'), and the changes due to collisions with the phonons (index 'ph').
The specific implementation of the individual terms on the right-hand side (rhs) of \cref{eq:f_els} is detailed in \cref{sec:coll}.

The thermal subsystems can be defined with two differential equations for the change of the internal energy densities $u$ of electrons and phonons, respectively:
\begin{subequations}\label{eq:ttmathem}
\begin{align}
    \dv{u\el}{t} &= - g\ep\,(T\el-T\ph) +~\!\left.\pdv{u\el}{t}\right|\elels\,, \label{eq:u_el}\\
    \dv{u\ph}{t} &= \phantom{-} g\ep\,(T\el-T\ph) + \left.\pdv{u\ph}{t}\right|\phels\,, \label{eq:u_ph}
\end{align}
\end{subequations}
where both thermal subsystems exchange their energy according to their temperature difference and 
the electron-phonon coupling parameter $g\ep$ as known from the standard TTM.
Additionally, both thermal subsystems are coupled to the laser-excited athermal system through the energy-transfer terms, which follow from the corresponding collision terms in \cref{eq:f_els} for each thermal subsystem $s\in \{\text{el, ph}\}$ as 

\begin{align}
    \left.\pdv{u_s}{t}\right|_{s-\text{el*}} &= \left. - \pdv{u\els}{t}\right|_{\text{el*}-s}\nonumber\\[0.5em]
    &=\, - \int \left.\pdv{f\els}{t}\right|_{\text{el*}-s}\,D(E)\,E\,\de\,. \label{eq:dudt}
\end{align}

Every integration in this manuscript goes over all considered energy states of the electrons. 
Furthermore, $D(E)$ denotes the electronic DOS, 
which is calculated from density functional theory, see Methods section.
Note that for the considered case of laser-excitation of the athermal electrons, the athermal electronic subsystem will lose energy, 
while the thermal electrons and thermal phonons will gain energy, thus the left and the right side of \cref{eq:dudt} for $s \in \{\text{el, ph}\}$ all yield positive values.

\begin{figure}
    \centering
    \includegraphics{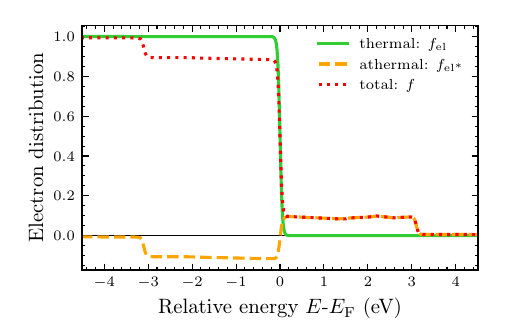}
    \caption{\textbf{Electron distributions in the AthEM model.} Example of the distributions of thermal (green, solid line) and athermal (orange, dashed line) electronic subsystems and the total electronic distribution (red, dotted line) in the AthEM model for aluminum after instantaneous excitation. The energy of the absorbed photons is \SI{3.1}{\eV}. The modelled energy density in this example is unusually high to clearly show the step-structure resulting from electron-hole generation.
    }
    \label{fig:athem}
\end{figure}

Within the AthEM, the athermal electron-hole pairs are initially excited by the laser pulse. 
Here, electrons are described as positive and holes as negative contributions to the athermal distribution $f\els(E)$. 
\Cref{fig:athem} shows exemplary curves for the thermal and athermal distribution, as they appear due to an instantaneous excitation of aluminum.
The sum of both distributions ("total" in \cref{fig:athem}), $f(E) = f\el(E) + f\els(E)$, gives the actual total laser-excited non-equilibrium electron distribution. 

The initial excitation, described by the first term on the rhs of \cref{eq:f_els}, creates an athermal distribution, $f\els$, that integrates to a total particle number, $n\els$, of zero since the number of holes, which contribute negatively to the distribution, and the number of excited electrons is equal, i.e. 
\begin{equation}
   \left.\pdv{n\els}{t}\right|_\text{laser}= \int \left.\pdv{f\els}{t}\right|_\text{laser} D(E)\,\de =0 \enspace. \label{eq:dndt_laser}
\end{equation}

However, depending on the construction of the particular collision terms, the second and third term on the rhs of \cref{eq:f_els} may not conserve particles within the athermal subsystem.
For instance, in many kinetic descriptions, a relaxation time approach (RTA) is applied with energy-dependent relaxation times, which does not satisfy conservation laws.
It is a great benefit of the AthEM, that the loss processes in the athermal subsystem can be balanced in the thermal electronic subsystem, as \cref{eq:dudt} strictly conserves energy.
An additional expression is also required to ensure total particle conservation. 
Any net particle loss of the athermal subsystem introduces particles in the thermal electron system,  
that is we set
\begin{align}
    \dv{n\el}{t} = &- \dv{n\els}{t}\label{eq:dndt}\\
    = &- \left.\pdv{n\els}{t}\right|\elsel 
    - \left.\pdv{n\els}{t}\right|\elsph \nonumber\\
    = &- \int \left(\left.\pdv{f\els}{t}\right|_{\text{el*-el}} 
      +  \left.\pdv{f\els}{t}\right|_{\text{el*-ph}}\right)\,D(E)\,\de \nonumber\enspace.
\end{align}
With this, the time-dependent equations that fully define AthEM are complete with equations (\ref{eq:f_els}), (\ref{eq:ttmathem}), (\ref{eq:dudt}) and (\ref{eq:dndt}).

%%%%%%%%%%%%%%%%%%%%%%%%%%%%%%%%%%%%%%%%%%%%%%%%%%%%%
\subsection{Collision terms in AthEM}\label{sec:coll}

In the following, we give the details of the applied collision terms in \cref{eq:f_els} entering the AthEM.

\subsubsection{Primary Electron Generation from Fermi's Golden Rule}

Absorption of a photon of energy~$\hbar \omega$ leads to the generation of excited electrons (primary electrons). The contribution that enters \cref{eq:f_els} due to laser excitation  is given by the sum of two processes
\begin{align}
    \left.\pdv{f\els(E)}{t}\right|\ii{laser} &= \left.\pdv{f\els(E)}{t}\right|_{h^+} + \left.\pdv{f\els(E)}{t}\right|_{e^-}\,. \label{eq:primary_electron_generation}
\end{align}
For given electron energy~$E$, electronic excitation can generate an electron~($e^-$) at $E_+=E+\hbar\omega$ and a hole~($h^+$) at energy $E$. Additionally, light excitation can create a hole at energy $E_-=E-\hbar\omega$ and an electron at $E$. Based on Fermi's golden rule, we can express the change in electronic distribution due to the first process as
\begin{align}
    \left.\pdv{f\els(E)}{t}\right|_{h^+} = \frac{2\pi V}{\hbar}\,D(E_+)\,|M_{E,E_+}|^2\,f(E)[1-f(E_+)]\,,\label{eq:holes_1}
\end{align}
where $f(E)$ is the total non-equilibrium distribution of the electrons at energy~$E$, as described above, and $V$ is the volume of the unit cell.
The factors containing the energy distribution $f$ describe the probability of finding available electrons at energy $E$ and free states at energy~$E_+$, respectively.
Furthermore, $|M_{E,E_+}|^2$ describes the transition matrix element, whose determination will be discussed below. 

Equivalently, the equation for the second process in \cref{eq:primary_electron_generation} can be written as
\begin{align}
    \left.\pdv{f\els(E)}{t}\right|_{e^-} = \frac{2\pi V}{\hbar}\,D(E_-)\,|M_{E,E_-}|^2\,f(E_-)[1-f(E)]\,.\label{eq:el_1}
\end{align}
In order to determine the two matrix elements $|M_{E,E_+}|^2$ and $|M_{E,E_-}|^2$, we assume that both are independent of energy. This approximation is motivated by the single-band approximation that we apply in the description of isotropic metals. This approximation is also imposed in our Boltzmann simulations. We note that this is not a fundamental restriction and, in the future, the primary electron generation can be expanded to angular-momentum-resolved representations where dipole transition rules can be more straightforwardly incorporated.
Since particle conservation has to be fulfilled in the two excitation processes, at any given time, the number of excited athermal holes equals the number of excited athermal electrons, satisfying
\begin{align}
\int \left.\pdv{f\els(E)}{t}\right|_{h^+}\,D(E)\,\de &= \int \left.\pdv{f\els(E)}{t}\right|_{e^-}\,D(E)\,\de . \label{eq:primary_electrons_particle_cons}
\end{align}
The total energy input at time $t$ can be expressed as 
\begin{align}
    s(t) &= \int  \left.\pdv{f\els(E,t)}{t}\right|\ii{laser}\,D(E)\,E\,\de\,,\label{eq:laser_power_density}
\end{align}
where \mbox{$s(t) = \alpha I(t)$} is the laser power density, which is given with the intensity of the laser pulse $I(t)$ and a constant absorption coefficient $\alpha$.
The requirements defined by \cref{eq:primary_electrons_particle_cons} and \cref{eq:laser_power_density} can be fulfilled by choosing the matrix elements at each point in time.
Further details can be found in the SI, section S4.

\subsubsection{Thermalization of Athermal Electrons} \label{sec:thermalization}

We use an approach motivated by a relaxation time approximation (RTA) to describe electron-electron thermalization, 
\begin{align}
    \left.\frac{\partial f\els(E)}{\partial t}\right|\elel = -\frac{f\els(E)}{\tau\elel(E)} + \frac{f\el(E)-f\rel(E)}{\tau\elel(E)} \enspace.\label{eq:athem_RTA}
\end{align}
Here, $f\rel(E) = f\rel(E, T\rel, \mu\rel)$ describes the thermalized ("relaxed") distribution, which is a Fermi-distribution at temperature $T\rel$ and chemical potential $\mu\rel$ with equivalent energy and particle density as the total non-equilibrium distribution~$f(E)=f\el(E)+f\els(E)$.
The first term on the rhs of \cref{eq:athem_RTA} describes recombination of electrons and holes.
Due to the change of the current thermal distribution $f\el$, also the athermal distribution has to change to capture the difference between $f\el$ and $f\rel$, which is described with the second term.
This improves the description of the thermalization compared to previous models~\cite{Carpene2006, Tsibidis2018, Uehlein2022}.
In the low perturbation regime, the second term is a very small correction and only becomes important for highly-excited systems.

We perform a two-dimensional root-finding to calculate a quasi electron temperature~$T\rel$ and quasi chemical potential~$\mu\rel$ for the relaxed distribution, using the fact that ~$f(E)=f\el(E)+f\els(E)$ and~$f\rel(E)$ contain the same energy and particle density. A numerically stable procedure for this is described in SI, section S5.

We use a non-constant energy-dependent relaxation time $\tau\elel(E)$.
It describes the lifetime of a single athermal electron colliding with thermal electrons 
and is given by Fermi liquid theory as the lifetime of an excited quasi particle~\cite{Mueller2013PRB, Pines, Kaveh1984}:
\begin{align}
    \tau\elel(E) &= \tau_0\frac{\mu^2}{(E-\mu)^2+(\pi k_BT_{\text{el}})^2}\,, \label{eq:tau_elel}\\
    \text{with}~~\tau_0 &= \frac{128}{\sqrt{3}\pi^2\omega\ii{p}}\,,
\end{align}
where $\omega\ii{p}$ is the plasma frequency of the material and $\mu$ the chemical potential of the thermal electrons.

We also use a simple RTA for the collision of athermal electrons with phonons. 
Therefore, we assume that due to this collision, the electron recombines with an athermal hole.
Thus, the relaxation can be written as 
\begin{align}
    \left.\pdv{f\els}{t}\right|\elph = - \frac{f\els}{\tau\elph} \,,\label{eq:RTA_el-ph}
\end{align}
where we apply the relaxation time of refs.~\citenum{
Carpene2006, Groeneveld1995},
\begin{align}
    \tau\elph &= \tau_\text{mfp}\frac{\hbar\omega}{k\ii{B}\theta\ii{D}}\, ,\label{eq:relaxtime_el-ph}
\end{align}
with the mean free path of the electrons $\tau_\text{mfp}$ and the Debye energy of phonons $k\ii{B} \theta\ii{D}$.
This relaxation time 
is based on the assumption that in each electron-phonon collision, occurring after $\tau_\text{mfp}$, the maximum phonon energy $k\ii{B} \theta\ii{D}$ is emitted by an electron with an excess energy caused by a laser excitation of $\hbar\omega$. 
It is a rough estimate yielding an upper limit for the energy relaxation rate due to electron-phonon scattering. 
For a constant relaxation time, the energy relaxation occurs at the same timescale as the relaxation of the electron distribution (see SI, section S6). 

%%%%%%%%%%%%%%%%%%%%%%%%%%%%%%%%%%%%%%%%%%%%%%%%%%%%%%%%%%%%%
\section{Discussion\label{sec:results}}
In the following, we will discuss the accuracy and benefits of AthEM by comparing its simulation results with eTTM and kinetic Boltzmann simulations for aluminum~(Al) and gold~(Au). AthEM and Boltzmann simulations use electronic DOSs for the two materials predicted from Density Functional Theory (DFT) calculations (see Methods section and Supplementary Figure S2). Aluminum is a material with a DOS that is similar to the free electron gas and was previously described well with eTTM and kinetic Boltzmann simulations in the RTA.
Gold represents a case where the high DOS of the occupied d-band becomes accessible for laser excitation at high photon energies, which is not accounted for in the eTTM.

\subsection{Primary Electron Generation} \label{sec:primelecgen}

\begin{figure}
    \centering
    \includegraphics{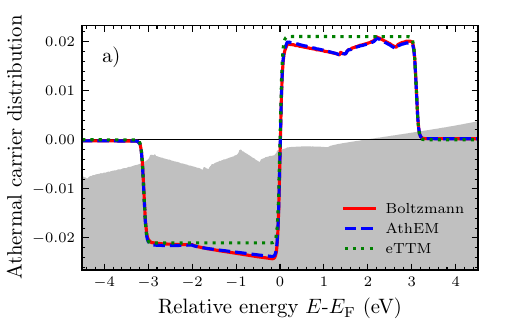}
    \includegraphics{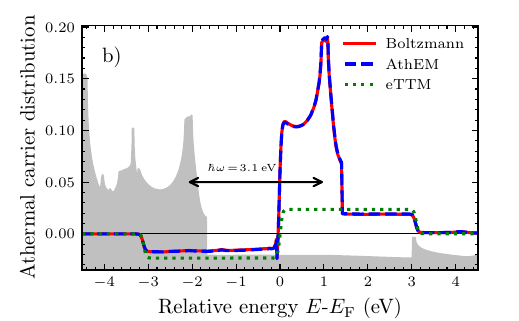}
    \caption{\textbf{Athermal carrier distributions as predicted by different models.} Comparison between kinetic Boltzmann simulations with full collision integral (red, solid line), AthEM (blue, dashed line) and eTTM (green, dotted line)~\cite{Uehlein2022}.
    Results are presented for aluminum, panel (a), and gold, panel (b). Distributions are shown immediately after a laser pulse with a photon energy of 3.1 eV. Thermalization processes are not considered. The DOS for Al and Au, respectively, is sketched in gray.
    }
    \label{fig:after_laser_step}
\end{figure}

We first examine the excitation process, whose description contains a significant improvement over the previous eTTM.
\Cref{fig:after_laser_step} shows eTTM, AthEM, and Boltzmann simulation results for athermal carrier distributions for aluminum and gold directly after a laser pulse of 25~fs duration with photon energy of 3.1~eV has ended. Thermalization is not considered in this comparison.
For the eTTM~\cite{Uehlein2022} and the AthEM, the athermal carrier distribution is given directly by the athermal electron subsystem (see \cref{fig:athem}).
We compare this to the difference between the excited distribution and the initial distribution of a calculation using a full Boltzmann collision integral for all electrons~\cite{Mueller2013PRB, Seibel2023}, 
which, in the absence of thermalization, equals the distribution of athermal electrons and holes.

In the case of aluminum (\Cref{fig:after_laser_step}~a), we find that all three methods provide closely matching athermal electron distributions. A steplike structure is established, where the width of the step is a result of the photon energy of \SI{3.1}{\eV} and the height of the step corresponds to the fluence of the laser pulse~\cite{Sun1994,Seibel2023}. 
That is, holes are nearly evenly generated within \SI{3.1}{\eV} below the Fermi level and electrons are nearly evenly generated within \SI{3.1}{\eV} above the Fermi level. Both AthEM and Boltzmann distributions match closely and capture subtleties in the distribution due to the slight variations in the DOS of aluminum, but the eTTM provides a fair approximation of the excited distribution. 

In contrast, in the case of gold (see \cref{fig:after_laser_step}~b), significant differences between the eTTM and the other models are apparent. The DOS of gold, predicted by GGA-PBE DFT calculations~\cite{perdew1996generalized}, shows the onset of the occupied d-bands at -1.6~eV below the Fermi level (gray-shaded area in \cref{fig:after_laser_step}~b). Therefore, we expect a high probability to excite electrons from this region if the photon energy is sufficiently high to reach those states. This results in an athermal distribution that reflects features of the DOS, shifted by the photon energy. As previously reported, this behavior is correctly captured in the Boltzmann model~\cite{Seibel2023}. The AthEM model results closely follow the Boltzmann simulation results. In contrast, the eTTM, where the athermal distribution is always described as a plain step-like function, does not capture such effects.

To emphasize the dependence of the DOS within $\pm\hbar\omega$ around the Fermi edge, we show in the SI the athermal distributions for gold after laser excitation with a photon energy of \SI{1.55}{\eV} (Supplementary Figure S3). The lower photon energy does not lead to direct d-band excitation, but two-photon excitation processes are still possible and lead to additional features. The fact that Boltzmann simulation and AthEM provide closely matching results in all cases gives some encouragement that the matrix element approximation in AthEM for the primary electron generation is reliable in the case of bulk metals.

As already demonstrated in \cref{fig:athem}, in AthEM, the total non-equilibrium distribution can be predicted at any point in time by adding the Fermi distribution of the thermal electrons to the athermal distribution.
The total electron distribution for the gold data shown in \cref{fig:after_laser_step} is shown in Supplementary Figure S4.
The shape of such distributions can be directly compared to experimental observations, for example, two-photon photoemission experiments~\cite{Fann1992b, Petek1997}. 

\subsection{Electron-electron thermalization}

\subsubsection{Advantage of splitting athermal and thermal electrons}

AthEM describes electron-electron scattering by transferring energy and particle density from the athermal to the thermal subsystem. As described in section \ref{sec:thermalization}, we impose the RTA to describe this process of electron thermalization. This is where AthEM's approach of splitting the electronic system into athermal and thermal subsystems provides a unique benefit. In kinetic Boltzmann simulations, electron-electron scattering is captured using explicit collision integrals, which comes at significant computational effort. Consequently, explicit electron-electron collision integrals are often replaced by thermalization in the RTA~\cite{DelFatti2000, Dubi2019, Sun1994, yadav_photocarrier_2019, bringuier_boltzmann_2019}, using either constant or energy-dependent relaxation times from Fermi-liquid theory. In this case, the electron-electron collision integral is replaced with an RTA,
\begin{align}
    \left.\pdv{f}{t}\right|_\text{el-el} = -\frac{f - f\rel}{\tau}. \label{eq:single_RTA}
\end{align}
Here, $f\rel$ is again a Fermi distribution with the same energy and particle density as $f$.
The temperature and chemical potential that can be attributed to this distribution $f\rel$ are called quasi temperature and quasi chemical potential of the actual distribution $f$.
In the cases where an energy-dependent $\tau=\tau(E)$ is used, \cref{eq:single_RTA} is not necessarily energy- or particle-conserving, as shown in SI, section S6. AthEM avoids this issue by having two electronic systems between which energy and particles can be transferred, with the energy lost from the non-equilibrium system entering the equilibrium one.
Particle conservation can be enforced in the same way.
When imposing the RTA with energy-dependent relaxation rates in the Boltzmann framework, there is only a single system and energy and particle conservation cannot be obeyed. 

\begin{figure}
    \centering
    \includegraphics{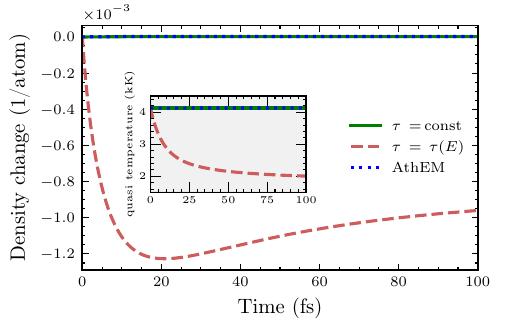}
    \caption{\textbf{Change in total electron density in aluminum.} Comparison between a Boltzmann simulation with RTA and constant relaxation time (green solid line), a Boltzmann simulation with RTA and energy-dependent relaxation time according to Fermi liquid theory (red dashed line), and an AthEM simulation with energy-dependent relaxation time according to Fermi liquid theory (blue dotted line), see \cref{eq:tau_elel}. Simulations consider an excited initial distribution which is allowed to thermalize. The inset shows the quasi-temperatures during the thermalization.} 
    \label{fig:relax}
\end{figure}

\Cref{fig:relax} shows the change in electron density during AthEM and Boltzmann equation simulations for electron-electron scattering in aluminum.
We initialize all calculations with the same non-equilibrium distribution, generated in advance with the Boltzmann collision term for laser excitation. Then, we trace the thermalization into a hot electron distribution with three different approaches, namely the AthEM, and the Boltzmann collision model with RTA and two different relaxation times. 
The Boltzmann equation using RTA and constant relaxation time maintains particle conservation throughout, whereas Boltzmann using RTA with the Fermi-liquid relaxation time, according to \cref{eq:tau_elel}, quickly drops the number of particles in the system. This not only questions the validity of the approach but also produces significantly different results to the true answer. This can be seen in the inset of \cref{fig:relax} which shows a significant reduction in the temperature by thousands of Kelvin. 
In contrast, AthEM with the same energy-dependent Fermi-liquid relaxation times maintains particle conservation throughout, and conserves the temperature in the closed system.

\begin{figure}
    \centering
    \includegraphics{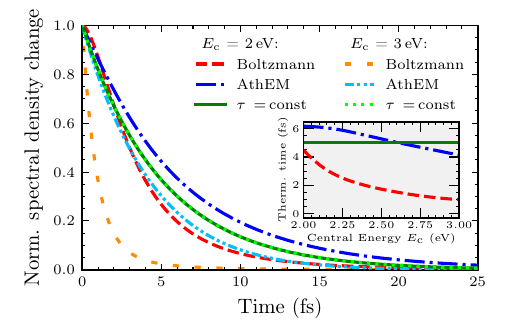}
    \caption{\textbf{Change in spectral densities in aluminum, normalized to the initial value.} Results are generated with AthEM (blue and light blue curves), Boltzmann equation with collision integral (red and orange curves), and Boltzmann equation with RTA and a constant lifetime for the electron-electron scattering (green and light green curves). The spectral densities are evaluated within an interval of $\Delta E = \SI{0.2}{\eV}$ centered around $E_c = \SI{2}{\eV}$ and $E_c = \SI{3}{\eV}$ above the Fermi edge. Simulations are initialized with an excited electron distribution. The inset shows the thermalization time evaluated by fitting an exponential decay as a function of the energy, $E_c$, at which the spectral density is evaluated.
    }
    \label{fig:spectral_all}
\end{figure}

The disadvantage of applying the Boltzmann model with constant relaxation time becomes clear, when observing energy-resolved (spectral) densities during the thermalization dynamics, which are of particular interest in comparison with experimental results~\cite{Kuehne2022}. 
The spectral density is defined as the number of particles within a window of width $\Delta E$ (\SI{0.2}{eV} in this case) and centered at $E_c$~\cite{Uehlein2022, Seibel2023}:
\begin{align}
    n_{E_\text{c}, \Delta E}(t) &= \int\displaylimits_{E_\text{c}-\Delta E/2}^{E_\text{c}+\Delta E/2} f(E, t)\,D(E)\,{\rm d}E\,.\label{eq:spec_dens}
\end{align}
The normalized spectral density change plotted in \cref{fig:spectral_all} is defined as the difference in density at time $t$ and the $t\rightarrow\infty$ limit:
\begin{align}
    \Delta\tilde{n}_{E_\text{c}, \Delta E}(t) &= \frac{n_{E_\text{c}, \Delta E}(t) - n_{E_\text{c}, \Delta E}(\infty)}{n_{E_\text{c}, \Delta E}(0)} . \label{eq:norm_spec}
\end{align}

In \cref{fig:spectral_all} we compare the spectral densities of a Boltzmann equation, with RTA and constant lifetime, and an AthEM simulation to a Boltzmann equation with full collision integrals.  
The full Boltzmann collision integral result yields a different behavior for $E_c = \SI{3}{\eV}$ and $E_c=\SI{2}{\eV}$.
The density change quickly decays at the higher energy and therefore exposes a fast relaxation whereas at \SI{2}{\eV} a slower decay is shown, evidencing that the relaxation at this energy is slower.
This is a result of two cooperative effects. Firstly, particles closer to the Fermi edge have a longer lifetime~\cite{Bauer2015}, this is the case when considering full collision integrals or Fermi-liquid theory.
Secondly, more particles will scatter into this energy window than into the \SI{3}{eV} window, maintaining a high number of particles over a longer duration~\cite{Seibel2023}. AthEM captures the same qualitative effect (spectral density at lower energies decays more slowly), however, the decay is slower than for the case calculated with Boltzmann collision integrals. 
The inset of \cref{fig:spectral_all} shows the thermalization time for a range of central energies, where AthEM consistently overestimates thermalization times compared to full scattering integrals in the Boltzmann equation. 
Importantly, when imposing a constant relaxation time within the Boltzmann equation, the thermalization time is independent of energy, lacking a key aspect of electron thermalization.

\subsubsection{Thermalization dynamics}

Having demonstrated the ability of AthEM to provide energy-dependent, as well as energy- and particle-conserving thermalization dynamics on the example of aluminum, we now discuss the details of the thermalization for gold. We again compare to the Boltzmann equation with full electron collision integrals and the previously proposed eTTM approach by Uehlein \textit{et al.}~\cite{Uehlein2022}.
As we will show here, the approach in AthEM is able to reproduce the overall thermalization behavior observed in the Boltzmann equation but sacrifices some fine detail in favor of numerical simplicity.

\begin{figure}
    \centering
    \includegraphics{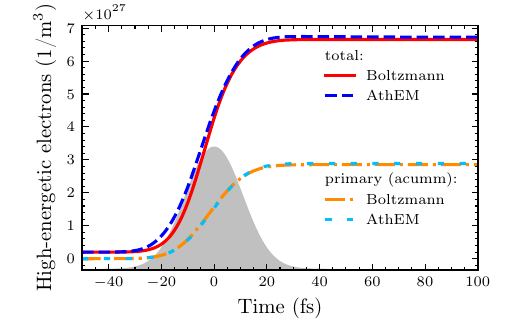}
    \caption{
    \textbf{Density of electrons in gold with energy above $E_F$ generated by laser excitation.} Shown are the total concentration of primary and secondary electrons simulated with Boltzmann (red, solid line) and AthEM (blue, short dashed line) and the sum of primary electrons generated by the laser source up to a point in time with Boltzmann (orange, dash-dotted line) and AthEM (light blue, long dashed line). The temporal laser profile is sketched in gray. 
    }
    \label{fig:prim_sec}
\end{figure}

As already discussed in \cref{sec:primelecgen}, the primary electron generation in AthEM closely agrees with Boltzmann simulations. \Cref{fig:prim_sec} shows that this also holds for high-energy electrons in the presence of electron thermalization. In the presence of electron-electron scattering, additional secondary electrons are generated above the Fermi edge through thermalization. For the total number of primary and secondary electrons, we also find generally good agreement between Boltzmann and AthEM (\cref{fig:prim_sec}).  The total electron generation differs only slightly during the start of the laser pulse, where AthEM starts generating high-energy electrons faster than Boltzmann. 

%%%%%%%%%%%%%%%%%%%%%%%%%%%%%%%%%%%%%%%%%%%%%%%%%%%%%%%%%%%%%%%%%%%%%%%%%

\begin{figure}
  \includegraphics{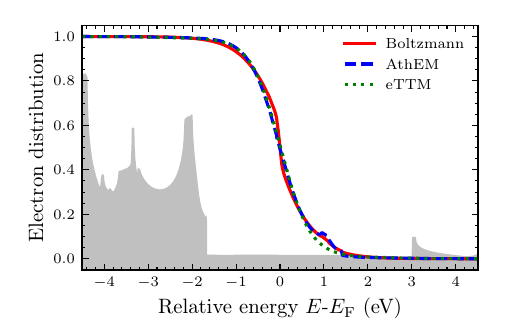}
  \caption{
  \textbf{Total electron distributions of gold at the peak of the laser pulse.} Simulations with Boltzmann (red, solid line), AthEM (blue, dashed line) and eTTM (green, dotted line) include electron thermalization. The shape of the DOS is shown in gray.
  }
  \label{fig:dist_max}
\end{figure}

Even if the total number of electrons above the Fermi edge is approximately equal for AthEM and Boltzmann, the distributions subtly differ. This is shown in \cref{fig:dist_max}, where the distributions for the time of the peak of the laser pulse, generated with the Boltzmann model, AthEM and eTTM, are compared.
While the distribution calculated with the eTTM shows no features at all, the AthEM distribution still shows an increased electron distribution at around \SI{1}{\eV} above the Fermi edge, which is a reflection of the high probability of excitation from the high-energy end of the d-bands at approximately $\SI{-2}{\eV}$. 
The Boltzmann model also shows this feature, but slightly less pronounced. 
Furthermore, directly around the Fermi edge, differences between the distributions of Boltzmann model and the other two models can be observed. %are visible.

\begin{figure}
  \includegraphics{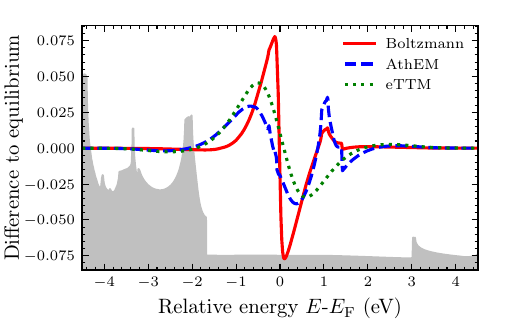}
  \caption{
  \textbf{Difference between simulated electron distribution and corresponding Fermi distribution $f\rel$.} Results are shown at the peak of the laser pulse for gold. Simulations with Boltzmann (red, solid line), AthEM (blue, dashed line) and eTTM (green, dotted line) include electron thermalization. The shape of the DOS is shown in gray.
  }
  \label{fig:ee_scatter_dis}
\end{figure}

These differences are visible more clearly in \cref{fig:ee_scatter_dis}. There, we show for all three models the difference between the non-equilibrium distribution and the corresponding thermal distribution, which is a Fermi distribution with the same energy and particle density, i.e. $f\rel$. 
The figure thus gives an impression of the non-equilibrium character of the distributions at the time of the peak of the laser pulse.
Compared to the final thermalized distribution, all distributions show an excess of electrons below the Fermi edge and a deficiency of electrons above it. 
Exceptions to this are features in energy regions above the Fermi edge caused by excitations from DOS peaks for the AthEM and Boltzmann model, where both models show an excess of electrons compared to the final distribution, while the eTTM neglects these features. AthEM appears to overestimate this effect compared to Boltzmann. Additionally, the Boltzmann model shows a higher deviation from equilibrium around the Fermi edge, which implies a slower thermalization rate close to the Fermi edge. 

A similar evaluation for aluminum is presented in the SI, section S8~C.

\subsection{Energy flow between subsystems}

\begin{figure}
    \centering
    \includegraphics{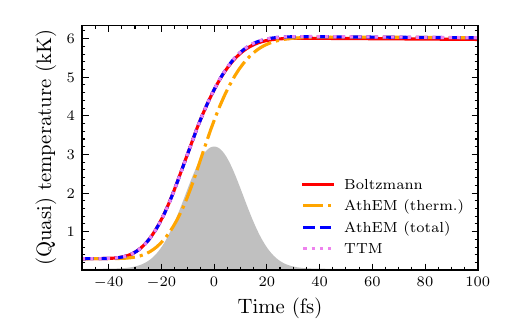}
    \caption{
    \textbf{Electron temperatures of gold extracted from Boltzmann model, AthEM, and TTM. }In case of the AthEM, the temperature of the thermal subsystem (orange, dash-dotted line) and the corresponding quasi-temperature of the total non-equilibrium distribution (blue, dashed line) is shown. These, and the quasi temperature for the Boltzmann model (red, solid line), are calculated by fitting to a corresponding Fermi distribution.
    The temporal laser profile is sketched in gray. 
    }
    \label{fig:temperatures}
\end{figure}
As a next step, we investigate the energy flow between the subsystems from an AthEM simulation for gold. Here, we also include the influence of the phonon subsystem.

The most succinct way to observe the total energy flow is by monitoring the subsystem temperatures.
\Cref{fig:temperatures} compares the electronic temperatures determined by a pure TTM simulation as well as the temperature of the thermal subsystem of the  AthEM, together with the electronic quasi temperature of all electrons considered in AthEM and the quasi temperature determined from the Boltzmann collision terms.
The pure TTM calculation is performed with the same energy input as the AthEM, and with the same electron-phonon coupling parameter as in the thermal part of the AthEM, $g_{ep}$ in \cref{eq:u_el}.
We use a temperature dependent parameter calculated using a full Boltzmann collision integral (see ref. \citenum{Held2025}).
Temperatures as predicted by all models agree well and exhibit a strong increase during laser irradiation.
The thermal subsystem of AthEM shows a temperature lag of a few femtoseconds compared to the quasi-temperatures and provides a clear indication of the thermalization time of the athermal subsystem.
At later times, 
the electron temperatures slowly 
decrease due to electron-phonon coupling. 

\begin{figure}
    \centering
    \includegraphics{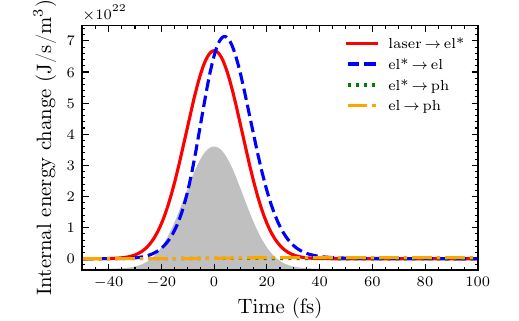}
    \caption{\textbf{Rate of change of internal energy for different processes during an AthEM simulation for gold.} 
    Shown are the energy transfer of the excitation process (red, solid line), from the athermal to the thermal electron system (blue, dashed line) and from both electron systems to the phonons (green, dotted and  orange, dash-dotted line).
    The temporal laser profile is sketched in gray. 
    }
    \label{fig:energy_change}
\end{figure}

\begin{figure}
    \centering
    \includegraphics{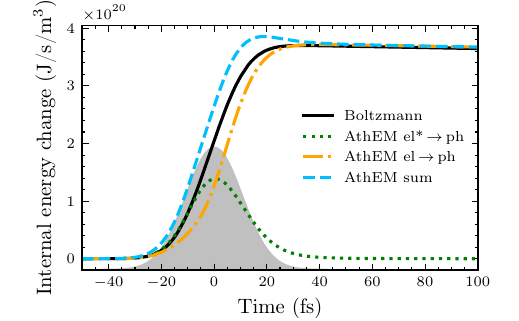}
    \caption{
    \textbf{Rate of change of internal energy due to electron-phonon collisions in gold.}
    Comparison between results from Boltzmann collision integral (black, solid line), both AthEM electron systems (green, dotted and orange dash-dotted line) and the sum of them (light blue, dashed line).
    The temporal laser profile is sketched in gray.
    }
    \label{fig:energy_change_ph}
\end{figure}

Mapping out individual energy transfer channels, \cref{fig:energy_change}  shows that the electron-electron thermalization from the athermal to the thermal electrons dominates the relaxation process. Electron thermalization occurs with only a few femtoseconds of delay after laser excitation. In contrast, the flow of energy from athermal as well as from thermal electrons to phonons is almost negligible due to the long relaxation time in the electron-phonon energy transfer. Thus, the athermal electrons rather decay into the thermal subsystem before significant energy transfer to the phononic subsystem takes place. This justifies the assumptions underlying \cref{eq:relaxtime_el-ph}.
On longer timescales, the thermal electron-phonon coupling process dominates. 

When we consider the energy flow into the phonon system separately, the influence of the athermal electrons on the phonons becomes more apparent. In \cref{fig:energy_change_ph}, we compare both electron-phonon processes captured in AthEM to the electron-phonon coupling of the Boltzmann model.
The Boltzmann relaxation behavior closely resembles that of the AthEM thermal system.
Overall, the electron-phonon energy flow is minor and slow, so the additional mechanism in AthEM does not have a large effect on any observable quantity. Therefore, also the simple RTA model for electron-phonon relaxation in \cref{eq:RTA_el-ph} is well justified for the case of gold.

\subsection{Comparison with experiment}\label{sec:exp}

\begin{figure}[t]
    \centering
    \includegraphics{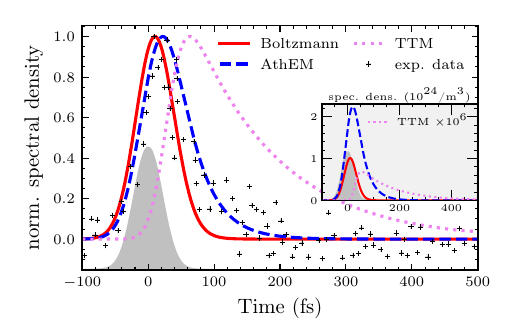}
    \caption{
    \textbf{Comparison of spectral densities for different models with experimental data.}
    Normalized spectral densities calculated with Boltzmann (red, solid line), AthEM (blue, dashed line) and TTM (pink, dotted line) are compared to the time-dependent signal measured by a photoemission experiment for gold~\cite{Kuehne2022}.
    The spectral densities are evaluated at \SI{1.2}{\eV} with an energy window of \SI{0.2}{\eV} and convoluted with a Gaussian with \SI{40}{\fs} FWHM, as suggested in ref. \citenum{Kuehne2022}.
    The inset shows the absolute spectral densities. Here, the result of the TTM is scaled by $10^6$. The temporal laser profile is sketched in gray.
    }
    \label{fig:spectral_exp}
\end{figure}

Electron spectral densities can be extracted out of non-equilibrium electron distributions~(see \cref{eq:spec_dens}), e.g. for comparison with photoemission experiments~\cite{Kuehne2022}. To allow a direct comparison, the laser parameters of the simulation have been set to a photon energy of \SI{1.55}{\eV}, FWHM of \SI{50}{\fs} and a lower fluence, resulting in peak electron temperatures below \SI{1000}{\kelvin}. 
We use \cref{eq:spec_dens} to calculate the spectral densities at a central energy of \SI{1.2}{\eV} above the Fermi edge, within an energy window of \SI{0.2}{\eV}. 
Furthermore, we convoluted the result with a Gaussian with a FWHM of \SI{40}{\fs} to consider the temporal resolution of the probe pulse, as suggested in the experimental paper.
\Cref{fig:spectral_exp} shows the convoluted spectral densities normalized to the peak value together with the experimental data~\cite{Kuehne2022}.
Additionally, the inset displays the non-normalized spectral densities.
To make the TTM data visible, it is scaled by a factor of \SI{1e6}{}.
Both AthEM and Boltzmann match the overall profile of the experiment well. 
However, the TTM completely fails to capture the spectral densities and exhibits slow electron decay, driven only by electron-phonon coupling. This provides evidence that the measured effect in the experiment results from athermal electrons, which cannot be described with a TTM.
The total values of the spectral densities vary between the models, with AthEM reaching more than twice the peak height of Boltzmann and both about six orders of magnitude greater than the TTM. Unfortunately, the experiment only provides a normalized spectral density, therefore, whether Boltzmann or AthEM predict more accurate spectral densities cannot be determined. 

\section{Conclusion}

In conclusion, we have developed a phenomenological model for light-matter interaction in metals, called AthEM, which accurately and efficiently captures non-equilibrium electron dynamics in materials. 
The model splits the electronic system into a thermal and an athermal subsystem. 
This development provides access to non-equilibrium electron dynamics in laser-excited materials over long time scales at significantly reduced computational effort compared to Boltzmann collision integral calculations. 
The model accurately captures primary electron generation, electron thermalization and electron-phonon relaxation upon laser excitation, which we show by comparing with calculations applying Boltzmann collision integrals and with experimental results.

The description of electronic dynamics in AthEM based on two subsystems enables the use of the RTA with energy-dependent relaxation rates from Fermi liquid theory. 
Contrary to Boltzmann equation simulations in the RTA with energy-dependent relaxation rates, AthEM satisfies energy and particle conservation. Although the RTA in AthEM slightly underestimates the relaxation rate of hot electrons as compared to full Boltzmann integral calculations, the model still provides physically interpretable thermalization dynamics and non-equilibrium distributions. 

The accuracy and computational efficiency of the AthEM make it an excellent platform for further extensions to consider, for example, incorporation of heat transport, including ballistic hot electron transport~\cite{Nenno2018}, electronic structure descriptions beyond the single band model~\cite{joao_atomistic_2023}, relaxation times beyond Fermi liquid theory~\cite{reinhard2015quantum}, or spin dynamics~\cite{Mueller2014PRB}. 
In general, the concept of a non-equilibrium electron subsystem presented here could be a viable approach to incorporate laser-induced non-equilibrium effects into other temperature-based models of spin, electron, and structure dynamics~\cite{Waldecker2016,Mueller2014PRB,Seibel2022,vanDriel1987}. 

\section{Methods}\label{sec:sim_details}

The numerical implementation of AthEM propagates the time evolution of the athermal distribution, as well as the temperature of the two thermal subsystems and the thermal electron particle density.
Propagating internal energy and particle density of the thermal electron subsystem or their temperature and particle density is an equivalent choice, as shown in SI, section S9.
We chose to follow the dynamics of the temperatures rather than internal energies, since the temperature is the essential quantity for electron-phonon relaxation appearing in the TTM (see \cref{eq:ttmathem}).
We transform the change in energy and particle density of the electrons directly into a change in temperature via the electronic distribution function (see SI, section S9).
For the phonons, we assume a constant heat capacity $c\ii{ph}$ connecting the change in energy density with a change in temperature.

The materials chosen to be under investigation are aluminum~(Al) and gold~(Au).
The parameters used to model them are given in \cref{tab:Parameters}.
The mean free path of electrons~$\tau\ii{mfp}$ is estimated according to refs.~\cite{Carpene2006, ashcroft} with data from refs.~\cite{Kittel, Halas1998}. The Fermi energy is the difference between the lower band edge and highest occupied state at \SI{0}{\kelvin}.
\setlength{\tabcolsep}{4pt}
\setlength\extrarowheight{2pt}
\begin{table}
    \centering
    \begin{tabular}{ccc}
    Parameter & Al & Au \\
    \hline
    Fermi energy (eV) & 11.15 & 9.94 \\
    \rowcolor[gray]{.95}[\tabcolsep]lattice constant (\AA) & 4.04 & 4.07 \\
    $\mathrm{\theta\ii{D}}$(K) & 428~\cite{Kittel} & 165~\cite{Kittel} \\
    \rowcolor[gray]{.95}[\tabcolsep]$c\ph$ (\SI{1e6}{\joule\per\kelvin\per\cubic\meter}) & 2.5~\cite{Uehlein2022} & 2.5~\cite{Uehlein2022}\\
    $\mathrm{\hbar\omega_p}$ (eV) & 14.98~\cite{Rakic1998} & 9.03~\cite{Rakic1998}\\
    \rowcolor[gray]{.95}[\tabcolsep]$\tau_\text{mfp}$ (fs) & 7.12 & 10.9\\
    \end{tabular}
    \caption{Material parameters for gold (Au) and aluminum~(Al) for the simulations.
    }\label{tab:Parameters}
\end{table}

Electronic DOSs that enter the simulations are taken from Density Functional Theory (DFT) calculations (Supplementary Figure S2) as implemented in FHI-aims~\cite{blum_ab_2009}, Version  230905. The calculations were performed using the default `tight` basis set, an energy cutoff of \SI{1e-6}{eV} and a force cutoff of \SI{1e-3}{eV\AA^{-1}}. The Perdew-Burke-Ernzerhof functional~\cite{perdew1996generalized} was utilized for all calculations, with a $k$-grid size of 12$\times$12$\times$12. These settings resulted in bulk lattice constants of \SI{4.07}{\AA} for gold and \SI{4.04}{\AA} for aluminum. The DOS were calculated using the tetrahedron method, with the $k$-grid interpolated by a factor of 20 along all directions and the energy domain ranging from -20 to 20 eV with the chemical potential set to \SI{0}{\eV}. 

To independently assess our results, we compare to Boltzmann model simulations similar to the ones reported in refs. \citenum{Mueller2013PRB, Seibel2023}. In this model, full Boltzmann collision integrals are used for all terms 
changing the distribution of the total electron system (see Supplementary Equation~(1)). The present work uses a modified form of the previously published model by using a random-$k$ approximation~\cite{Knorren2000, Penn1985}. 
Further details can be found in SI, section S7.

We use a Gaussian laser pulse for the calculations reported in section \ref{sec:results}. 
The parameters are given in SI, section S7. We used a pulse with a full width at half maximum (FWHM) of \SI{25}{\fs} and a photon energy of \SI{3.1}{\eV} (set A and B in Supplementary Table S1).
To compare to experiment, we used a different set of laser parameters in \cref{sec:exp}. Those parameters are provided in Supplementary Table~S1 (set C).

Furthermore, we use a temperature-dependent electron-phonon coupling parameter $g_\text{ep} = g_\text{ep}(T\el, T\ph)$ for the coupling between the thermal electron system and the phonons, as described in ref.~\citenum{Held2025}. It is calculated using a full Boltzmann collision integral and a plane-wave matrix element for all electrons of the material. 

\section*{Acknowledgements}

R.J.M. acknowledges support through the UKRI Future Leaders Fellowship programme (MR/X023109/1), and a UKRI frontier research grant (EP/X014088/1). M.U., C.S., S.T.W. and B.R. acknowledge support through the Deutsche Forschungsgemeinschaft (DFG, German Research Foundation) - TRR 173 - 268565370 Spin+X (project no. A08).
High-performance computing resources were provided via the Scientific Computing Research Technology Platform of the University of Warwick and the EPSRC-funded HPC Midlands+ computing centre for access to Sulis (EP/P020232/1).
We appreciate the Allianz für Hochleistungsrechnen Rheinland-Pfalz for providing computing resources through project STREMON on the Elwetritsch high-performance computing cluster.

\section*{Data availability}
The data that support the findings of this study are available from the corresponding authors upon reasonable request.
 
\section*{Author contributions}
M.U. and H.S. developed the model and processed the data.
M.U., H.S., R.J.M. and B.R. wrote the manuscript.
C.S., T.H. and S.T.W. contributed to discussion during model development, interpretation of results, and revision of the manuscript.
R.J.M. and B.R. conceptualised the research and provided resources.

\section*{Additional information}
\textbf{Supplementary information} is available for this paper below containing a more detailed description of the computational methods, along with some theoretical fundamentals and supplementary figures. It includes refs. \cite{Mueller2011, Brouwer2014, Ndione2022}.

	\bibliography{bibfile/Main.bib}

\end{document}

\title{Supplementary Information for \\"Non-equilibrium electron dynamics  in metals modelled accurately and efficiently"}
\author{M. \surname{Uehlein}}
	\affiliation{Department of Physics and Research Center OPTIMAS, RPTU Kaiserslautern-Landau, 67663 Kaiserslautern, Germany}
 
	\author{H. T. \surname{Snowden}}
	\affiliation{Department of Chemistry, University of Warwick, CV4 7AL Coventry, United Kingdom}
        \author{C. \surname{Seibel}}	
	\affiliation{Department of Physics and Research Center OPTIMAS, RPTU Kaiserslautern-Landau, 67663 Kaiserslautern, Germany}

 \author{T. \surname{Held}}	
	\affiliation{Department of Physics and Research Center OPTIMAS, RPTU Kaiserslautern-Landau, 67663 Kaiserslautern, Germany}
 
        \author{S. T. \surname{Weber}}	
	\affiliation{Department of Physics and Research Center OPTIMAS, RPTU Kaiserslautern-Landau, 67663 Kaiserslautern, Germany}
 
 \author{R. J. \surname{Maurer}}
 \email{r.maurer@warwick.ac.uk}
 \affiliation{Department of Chemistry, University of Warwick, CV4 7AL Coventry, United Kingdom}
 \affiliation{Department of Physics, University of Warwick, CV4 7AL Coventry, United Kingdom}
 
	\author{B. \surname{Rethfeld}}
 \email{rethfeld@rptu.de}
	\affiliation{Department of Physics and Research Center OPTIMAS, RPTU Kaiserslautern-Landau, 67663 Kaiserslautern, Germany}
    
\date{\today}

\maketitle

%%%%%%%%%%%%%%%%%%%%%%%%%%%%%%%%%%%%%%%%%%%%%%%%%%%%%
\section{From Boltzmann collision terms to two-temperature model}\label{sec:blg2ttm}

The Boltzmann equation for an ensemble of particles in a homogeneous medium and without external forces reduces to the temporal change of the considered particle distribution. 
It is often applied to describe the changes of the electrons' distribution~$f$ during and after interaction with an ultrashort laser pulse, commonly including a term for the laser excitation, the electron-electron interaction and the electron-phonon interaction \cite{Rethfeld2002,Pietanza2004,Mueller2013PRB}. 
The latter process also changes the phononic distribution~$g$. 
The equation system can thus formally be written as, 
\begin{subequations}\label{eq:blg allg}
\begin{align}
    \dv{f}{t} = \pdv{f}{t}& = \left.\pdv{f}{t}\right|\ii{el-el} + \left.\pdv{f}{t}\right|\ii{el-ph}  + \left.\pdv{f}{t}\right|\ii{laser} \,,\label{eq:blg el}\\
    \dv{g}{t} = \pdv{g}{t} &= \left.\pdv{g}{t}\right|\ii{ph-el}\,,\label{eq:blg ph}
\end{align}
\end{subequations}
where the details of the individual collision terms can differ in different implementations of such a description. 

The first moment of the distribution function $f =f(E,t)$, depending on energy $E$ and time $t$, is an integral determining the energy density of the electrons 
\begin{equation}
    u\el(t)  = \int f (E,t)\, D(E)\, E\, \de\,, \\
\end{equation}
with the density of states $D(E)$, depending on energy but not on time.
Every integration, unless otherwise specified, goes over all considered energy states of the electrons.
The temporal change of the energy density can thus be determined with \cref{eq:blg el} as 
\begin{equation}
    \pdv{u\el}{t} = \int \pdv{f}{t}\,D(E)\,E\,\de\,,\label{eq:dudt total}\\
\end{equation}
which can be analogously defined for the phononic energy density $u\ph$ by integrating the distribution $g(E\ph,t)$ over the phonons' energies $E\ph$. 
Note that the energy changes due to electron-phonon interaction differ only by their sign, i.e.
\begin{equation}\label{eq:encnsv pe}
    \left.\pdv{u\ph}{t}\right|\ii{ph-el} = - \left.\pdv{u\el}{t}\right|\ii{el-ph}   \,, \\
\end{equation}
which expresses the energy conservation between the two considered subsystems.

In many cases, the details of the distribution functions are not relevant and the interaction between electrons and phonons can be described as if the electrons were in thermal equilibrium, i.e. following a Fermi-Dirac distribution at temperature $T\el$, while the phonons are in a Bose-Einstein distribution at temperature $T\ph$. 
The laser will initially heat the electrons, see \cref{eq:blg el}, thus usually and in particular for an optical laser, $T\el > T\ph$.
Both temperatures may be changing in time. 
The integral of \cref{eq:blg el} according to \cref{eq:dudt total} results in an energy balance, and the right-hand side can be evaluated separately for the different collision terms considered in \cref{eq:blg allg}. 
Here, in case of a Fermi-Dirac distribution, the electron-electron collision term $\left(\partial f/\partial t\right)|\ii{el-el}$ vanishes. 
The second term, given by electron-phonon collisions, can be reduced to an expression depending on the temperature difference, $(T\el-T\ph)$, as shown in refs.~\citenum{Kaganov1957,Anisimov1974}. 
The laser excitation, last term on the right-hand side of \cref{eq:blg el}, introduces a certain power density $s(t)$ into the electronic subsystem. 
As for the phononic subsystem, the same consideration holds, with the same energy-exchange term according to \cref{eq:encnsv pe}.
We therefore arrive at the equation system 
\begin{subequations}\label{eq:ttmplain}
    \begin{align}
        \dv{u\el}{t} = c\el\,\pdv{T\el}{t} &= - g_{ep}\,(T\el-T\ph) + s(t)\,, \label{eq:u_el}\\
        \dv{u\ph}{t} = c\ph\,\pdv{T\ph}{t} &= \phantom{-} g_{ep}\,(T\el-T\ph) \,, \label{eq:u_ph}
\end{align}
\end{subequations}
which  represents the well-known two-temperature model (TTM)~\cite{Anisimov1974}, for the case of a uniformly heated material. 
Here, $c\el$ and $c\ph$ are the specific heats of electrons and phonons, respectively, while $g_{ep}$ is the electron-phonon coupling parameter.

%%%%%%%%%%%%%%%%%%%%%%%%%%%%%%%%%%%%%%%%%%%%%%%%%%%%%
\section{Considering two electron systems and one phonon system}\label{sec:twoel}

When the material contains a second electronic system, it can be considered in an equal manner,  and additionally both electronic systems interact with each other. 
This can be formally written as,
\begin{subequations}\label{eq:blg twoel}
\begin{align}
    \pdv{f\els}{t}& = \left.\pdv{f\els}{t}\right|\ii{el*-el*} + \left.\pdv{f\els}{t}\right|\elel + \left.\pdv{f\els}{t}\right|\elph  + \left.\pdv{f\els}{t}\right|\ii{laser} \,,\label{eq:blg el* of 2 el}\\
    \pdv{f\el}{t}& = \left.\pdv{f\el}{t}\right|\ii{el-el} + \left.\pdv{f\el}{t}\right|\ii{el-el*} + \left.\pdv{f\el}{t}\right|\ii{el-ph}  + \left.\pdv{f\el}{t}\right|\ii{laser} \,,\label{eq:blg el of 2 el}\\
    \pdv{g}{t} &= \left.\pdv{g}{t}\right|\ii{ph-el} + \left.\pdv{g}{t}\right|\ii{ph-el*} \,,\label{eq:blg ph of 2 el}
\end{align}
\end{subequations}
where the subscripts 'el' and 'el*' denote the two distinct electronic subsystems.
Generally, different cases with two electronic subsystems can be imagined, e.g. a ferromagnet~\cite{Mueller2011,Mueller2013PRL}, a dielectric material~\cite{Brouwer2014,Brouwer2017}, or a noble metal with separated $d$- and $sp$-electrons~\cite{Ndione2022b}. 

Also in such cases, the energy-transfer processes are energy conserving, thus additionally to \cref{eq:encnsv pe}, we have
\begin{equation}\label{eq:encnsv pe*}
   \left.\pdv{u\ph}{t}\right|\ii{ph-el*} = - \left.\pdv{u\els}{t}\right|\elph \,,
\end{equation}
moreover, within the two electronic subsystems not only the total energy density, $u_e + u_{e^*}$, but also the total electronic particle density, $n_e+n_{e^*}$, is conserved, thus
\begin{subequations}\label{eq:encnsv pcncv ee}
\begin{align}
   \left.\pdv{u\el}{t}\right|\ii{el-el*} &= - \left.\pdv{u\els}{t}\right|\elel \,,\label{eq:encnsv ee}\\
   \left.\pdv{n\el}{t}\right|\ii{el-el*} &= - \left.\pdv{n\els}{t}\right|\elel \,,\label{eq:pcncv ee}
\end{align}
\end{subequations}
where the particle density is given as the zeroth moment of the distribution,
\begin{equation}
    n\ii{el/el*}(t)  = \int f\ii{el/el*} (E,t)\, D(E)\, \de\,. \\
\end{equation}

Specifically, for the AthEM, we consider the system 'el*' as the laser-excited {\em athermal} electronic subsystem 
and the system 'el' as a Fermi-distributed {\em thermal} electronic subsystem.
The laser interacts solely with the subsystem of athermal electrons, introducing excited 
electrons above Fermi energy and holes below Fermi energy. 
The temperature of the thermal electrons is increased by energy transfer from the athermal electrons and decreased by interaction with the phononic subsystem. 

Parts of the energy balance can be formulated analogously to the two-temperature model (TTM), see \cref{eq:ttmplain}, namely the energy density of the thermal electrons and the phonons. 
However, the description of the dynamics of the athermal electrons is kept directly via the temporal evolution of the distribution according to \cref{eq:blg el* of 2 el}. Note that we neglect the interaction of athermal electrons with each other, thus we set $\left(\partial f\els/\partial t\right)|\ii{el*-el*} = 0$.

Energy conservation between all subsystems is ensured by applying eqs.~(\ref{eq:encnsv pe*}) and (\ref{eq:encnsv ee}), while the electronic particle density is maintained through \cref{eq:pcncv ee}.
Note that the initial excitation of the athermal electron system does not change the density of thermal electrons. 
It is, however, nonconstant in the latter course of electron thermalization due to the specific construction of the electron-electron interaction within the athermal subsystem, see also section~\ref{sec:relax}.

%%%%%%%%%%%%%%%%%%%%%%%%%%%%%%%%%%%%%%%%%%%%%%%%%%%%%

\section{Violation of particle conservation in the eTTM}
\begin{figure}
    \centering
    \includegraphics{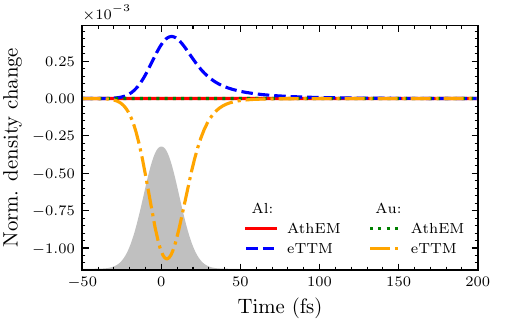}
    \caption{
    Change of electron density as compared to the start value (see \cref{eq:particles}).
    The shape of the laser pulse is sketched in gray and parameter sets \cref{tab_laser}~A and B are used.
    During the laser pulse, the eTTM results in a particle increase for aluminum and a particle decrease for gold.
    The AthEM conserves particles within the given accuracy.
    }
    \label{fig:particles}
\end{figure}

In \cref{fig:particles}, we show the electrons densities $n(t)$ calculated with AthEM and eTTM for aluminum and gold.
We normalize the particle density to their initial value and display the change of this value:
\begin{align}
    \tilde{n}(t) = \frac{n(t)}{n(t=0)}-1\label{eq:particles}
\end{align}
The AthEM maintains for both materials a detailed balance of the particle density.
In contrast, the eTTM differs from detailed balance.
We observe a maximum increase in particles of \SI{1.2e-3}{} electrons per atom in aluminum, while a maximum decrease of \SI{11.8e-3}{} electrons per atom is calculated for gold.
This violation of particle conservation exists during the entire time in which athermal electrons exist.
Both the direction and the strength of the violation can be traced back to the shape of the DOS.
Gold, which has a large increase below the Fermi edge, shows a large loss of electrons, because the eTTM underestimates the amount of athermal electrons that should be excited.
In contrast to this, aluminum, which has above the Fermi edge a slightly higher DOS than below, a small increase in the number of electrons is observed.
Please note, that larger excitations lead to a larger violation of the particle conservation in the eTTM.

%%%%%%%%%%%%%%%%%%%%%%%%%%%%%%%%%%%%%%%%%%%%%%%%%%%%%

\section{Primary electron generation}

The transition matrix elements~$|M_{E,E'}|^2$ of the excitation describe, in first order, the coupling of electronic states, $E$, $E'$, in the material with the electromagnetic dipole field induced by the laser:
\begin{align}
    M_{E,E'} = -\frac{e}{mc}\braket{\mathbf{k}_E|\mathbf{A}\cdot \mathbf{p}|\mathbf{k}_{E'}} ,
\end{align}
where $\mathbf{A}$ is the electromagnetic vector potential and $\mathbf{p}$ is the momentum operator. 
The laser pulse has a given power density~$s(t)$ and temporal envelope, which define the amplitude, $A_0$ and frequencies described by the vector potential~$\mathbf{A}$. Herein, we treat isotropic metals in a single-band approximation and assume that the transition matrix elements are energy-independent. Their magnitude is given by the field amplitude which is directly related to the power density $s(t)$ and the requirement that excitations conserve the number of particles, generating holes and electrons in equal numbers.

To achieve this, we first introduce two energy-independent transition rates
\begin{subequations}
\begin{align}
    \delta_{h^+} &= \frac{2\pi V}{\hbar}\,|M_{E,E_+}|^2\,, \\
    \delta_{e^-} &= \frac{2\pi V}{\hbar}\,|M_{E,E_-}|^2\,.
\end{align}
\end{subequations}
With this, the equations for the change of the distribution according to laser excitation can be written as
\begin{subequations}
\begin{align}
    \left.\pdv{f\els(E)}{t}\right|_{h^+} &= \delta_{h^+}\,D(E_+)\,\mathcal{F}_{h^+}(E)\,,\label{eq:holes_2}\\
    \left.\pdv{f\els(E)}{t}\right|_{e^-} &= \delta_{e^-}\,D(E_-)\,\mathcal{F}_{e^-}(E)\,,\label{eq:el_2}
\end{align}
\end{subequations}
with the definitions
\begin{subequations}
\begin{align}
    \mathcal{F}_{h^+}(E) &= f(E)[1-f(E_+)]\,,\\
    \mathcal{F}_{e^-}(E) &= f(E_-)[1-f(E)]\,.
\end{align}
\end{subequations}
Due to particle conservation, we set
\begin{subequations}
\begin{align}
    \int \left.\pdv{f\els(E)}{t}\right|_{h^+}\,D(E)\,\de &= \int \left.\pdv{f\els(E)}{t}\right|_{e^-}\,D(E)\,\de\\
    \Leftrightarrow~~~\delta_{h^+} \int D(E_+)\,\mathcal{F}_{h^+}(E)\,D(E)\,\de &= \delta_{e^-} \int D(E_-)\,\mathcal{F}_{e^-}(E)\,D(E)\,\de\,.\label{eq:delta_n}
\end{align}
\end{subequations}
To match the laser power density~$s(t)$ of the laser pulse, we set
\begin{subequations}
\begin{align}
    s(t) &= \int \left.\pdv{f\els(E)}{t}\right|_{h^+}\,D(E)\,E\,\de
         + \int \left.\pdv{f\els(E)}{t}\right|_{e^-}\,D(E)\,E\,\de\\
    \Leftrightarrow~~~s(t) &= \delta_{h^+} \int D(E_+)\,\mathcal{F}_{h^+}(E)\,D(E)\,E\,\de        
    + \delta_{e^-} \int D(E_-)\,\mathcal{F}_{e^-}(E)\,D(E)\,E\,\de\,.\label{eq:delta_u}
\end{align}
\end{subequations}
After evaluating the integrals at each point in time, we solve the system of equations~(\ref{eq:delta_n}) and~(\ref{eq:delta_u}) to determine ~$\delta_{h^+}$ and~$\delta_{e^-}$.
This results in the contribution due to laser excitation that enters the differential equation system:
\begin{align}
    \left.\pdv{f\els(E)}{t}\right|\ii{laser} &= \left.\pdv{f\els(E)}{t}\right|_{h^+} + \left.\pdv{f\els(E)}{t}\right|_{e^-}\\
    &= \delta_{h^+}\,D(E_+)\,\mathcal{F}_{h^+}(E) + \delta_{e^-}\,D(E_-)\,\mathcal{F}_{e^-}(E)\,.\nonumber
\end{align}

%%%%%%%%%%%%%%%%%%%%%%%%%%%%%%%%%%%%%%%%%%%%%%%%%%
\section{Numerically stable two-dimensional root-finding for temperature and chemical potential}
We perform a two-dimensional root-finding to calculate a quasi-temperature and quasi-chemical potential out of an arbitrary non-equilibrium electron distribution~$f$.
These quantities define a Fermi distribution~$f\rel$ with the same energy and particle density as the non-equilibrium distribution.
The deviations of a non-equilibrium distribution from a Fermi distribution are usually close to the Fermi edge, both, above and below the edge.
In the low perturbation regime, this causes round-off errors at energies below the Fermi edge.

To increase the numerical stability, accuracy and velocity we introduce a root-finding energy density $u\iroot$ and a root-finding particle density $n\iroot$ by transforming the integrals that calculate the energy and particle density out of an arbitrary distribution~$f'$:
\begin{subequations}
\begin{align}
    u[f'] &= \int_0^\infty f'(E)\,D(E)\,E\,\de \\
    &= \underbrace{\int_{0}^{E\iF} (f'(E)-1)\,D(E)\,E\,\de+ \int_{E\iF}^\infty f'(E)\,D(E)\,E\,\de}_{=u\iroot[f']}\nonumber + \int_{0}^{E\iF} D(E)\,E\,\de \\
    &= u\iroot[f'] + u_0\,,\nonumber\\
    n[f'] &= \int_0^\infty f'(E)\,D(E)\,\de = n\iroot[f'] + n_0\,.
\end{align}
\end{subequations}
Here, we set the energy $E$ to be zero at the lowest considered energy state.
We split the integration at the Fermi energy~$E\iF$, which is the difference between the lower band edge and the energy of the highest occupied state at \SI{0}{\kelvin}.
The calculation of $u\iroot$ and $n\iroot$ is numerically more accurate, than the calculation of the original integrals, because it prevents the round-off errors in the original integrals below the Fermi edge. 
The integrals of $u_0$ and $n_0$ are independent of the distribution.
Therefore, the root-finding can be performed by using the root-finding energy and particle density
\begin{subequations}
\begin{align}
    u\iroot[f] &= u\iroot[f\rel]\,, \\
    n\iroot[f] &= n\iroot[f\rel]\,.
\end{align}
\end{subequations}

%%%%%%%%%%%%%%%%%%%%%%%%%%%%%%%%%%%%%%%%%%%%%%%%%%%%%
\section{Limits of the relaxation time approach}\label{sec:relax}

The relaxation time approach
\begin{align}
    \frac{\partial f(E)}{\partial t} = \frac{f\rel(E)-f(E)}{\tau}
\end{align}
was initially formulated using a constant relaxation time $\tau$.
Here, the relaxed, or thermalized distribution $f\rel(E)$ is a Fermi distribution with the same energy and particle content as the current distribution $f(E)$.

It is straightforward to show that for a constant relaxation time the particle density (0th moment of the distribution, $i=0$) and the internal energy density (first moment of the distribution, $i=1$) are conserved. 
For the remainder of a discussion, we use the index $i=0,1$ to represent the number of particles ($x_0$) and internal energy ($x_1$).
The change of energy or particle density over time is then given by
\begin{subequations}
\begin{align}
    \frac{\partial x_i}{\partial t} &= \int \frac{\partial f(E)}{\partial t}D(E)E^i \, \de \\
    &= \int \frac{f\rel(E)-f(E)}{\tau} D(E) E^i \, \de \\
    &= \frac{1}{\tau}\int (f\rel(E)-f(E)) D(E) E^i \, \de \\
    &= \frac{1}{\tau}\left(\int f\rel(E) D(E) E^i \, \de-\int f(E) D(E) E^i \, \de\right)\\
    &= \frac{1}{\tau} (x_i - x_i) = 0\,.
\end{align}
\end{subequations}

Due to the definition of $f\rel$, the change in internal energy density or particles density must be zero.

However, for an energy-dependent relaxation time, $\tau(E)$ cannot be removed from the integral, so the final expression is therefore,
\begin{align}
    \frac{\partial x_i}{\partial t} &= \int \frac{f\rel(E)}{\tau(E)} D(E) E^i \, \de -\int \frac{f(E)}{\tau(E)} D(E) E^i \, \de\,.
\end{align}
This does not inherently cancel and could only do so with very specific forms of $\tau(E)$. 
Therefore, in general, a relaxation time approach with an energy-dependent relaxation time does not conserve energy and particle densities.

Once one introduces a second electronic system, they can correct for the lack of particle and energy conservation by transferring the negation of each quantity into the second system. This is what is performed in AthEM by defining the transfer of internal energy and particle density between the athermal and thermal electronic system as described in the main manuscript. 
This method utilizes and corrects the energy and particle conservation by transferring the difference from the athermal to the thermal system, therefore the total energy and particle conservation is constant.

%%%%%%%%%%%%%%%%%%%%%%%%%%%%%%%%%%%%%%%%%%%%%%%%%%%%%

\section{Further Simulation Details}
\begin{figure}
    \centering
    \includegraphics{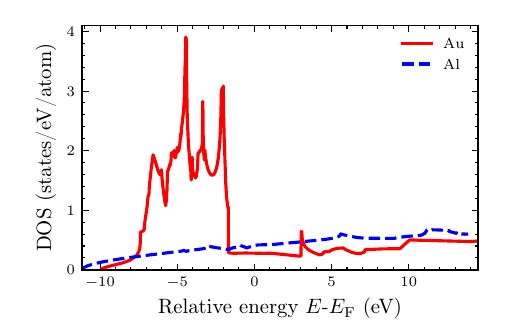}
    \caption{Density of states~(DOS) of gold~(Au) and aluminum~(Al) calculated using Density Functional Theory. Aluminum resembles a free electron gas DOS, while gold shows high d-peaks below the Fermi edge.}
    \label{fig:dos}
\end{figure}

We demonstrate our model by using aluminum (Al) and gold (Au) as example materials. The DOS of the two materials used for the calculations is given in \cref{fig:dos}. Details on how they are calculated can be found in the Methods section of the main manuscript. 

The parameters, that define the material, are listed in table I of the main manuscript.
\setlength{\tabcolsep}{4pt}
\setlength\extrarowheight{2pt}
\begin{table}
    \centering
    \begin{tabular}{ccccc}
    Parameter & A & B & C & D \\
    \hline
    material& Al & Au & Au & Au \\
    \rowcolor[gray]{.95}[\tabcolsep]full width at half maximum (fs) & 25 & 25 & 50 & 25  \\
    photon energies $\hbar\omega$ (eV) & 3.1 & 3.1 & 1.55 & 1.55\\ 
    \rowcolor[gray]{.95}[\tabcolsep]absorbed energy density (\SI{}{\joule\per\meter\cubed})\hspace{.2cm} & \SI{8.1e8}{}& \SI{1.7e9}{} & \SI{1.0e7}{} & \SI{3.3e9}{}
    \end{tabular}
    \caption{
    Four different sets of laser parameter-material combinations. 
    }\label{tab_laser}
\end{table}

The four different laser configurations used throughout the paper are given in \cref{tab_laser}. Sets A and B are used in all results sections up to section 'Comparison with experiment', where set C is used. Set D is used in SI, section \ref{sec:small}.

Our AthEM results are compared to simulations with a Boltzmann model.
For each collision process, the change of the distribution for electrons with momentum $k$ is of the form
\begin{equation}
    \left.\pdv{f(k)}{t}\right|_\text{coll} = \sum_{\text{all }k'} M^2(k, k')\times\mathcal{F}(f(k), f(k'), f(k\pm k'))\times\delta(E(k), E(k\pm k')).
\end{equation}
Here, we consider all collision partners~$k'$, while assuming an isotropic momentum space. 
The quantum mechanical transition probability~$M^2$ is calculated using a screened Coulomb potential and a plane wave approach. The collision functional~$\mathcal{F}$ determines the probability for a transition based on the Pauli exclusion principle. The model is described in more detail in refs.~\citenum{Mueller2013PRB, Seibel2023}. However, here we use the random-$k$ approximation~\cite{Knorren2000, Penn1985} to calculate the electron-electron collisions.

%%%%%%%%%%%%%%%%%%%%%%%%%%%%%%%%%%%%%%%%%%%%%%%%%%%%%
\section{Details of electronic distribution}

%------------------------------------
\subsection{Excitation of gold with photon energy not reaching d-bands}\label{sec:small}
\begin{figure}
    \centering
    \includegraphics{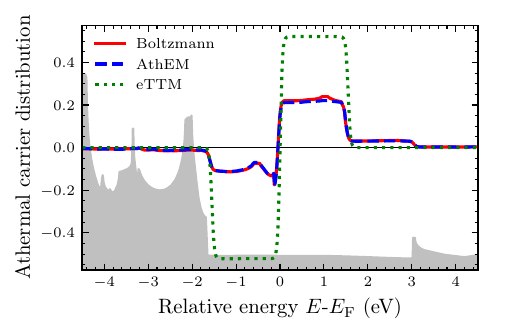}
    \caption{Athermal carrier distributions as predicted by AthEM, eTTM~\cite{Uehlein2022}, and kinetic Boltzmann simulations with full collision integral for Au. Distributions are shown immediately after a laser pulse with a photon energy of \SI{1.55}{\eV}. Thermalization processes are not considered. The DOS for Au is sketched in gray and parameter set \cref{tab_laser}~D is used.
    }
    \label{fig:after_laser_au_800}
\end{figure}
In \cref{fig:after_laser_au_800}, we show the distributions in gold after laser excitation with a photon energy of \SI{1.55}{\eV}. All thermalization processes are neglected. Due to the photon energy of \SI{1.55}{\eV}, no electrons can be excited directly from the d-band.
The material is excited with a high fluence, therefore multi-photon absorption can take place, giving rise to finite carrier distributions above and below the one-photon-energy window above and below the Fermi energy.
This is not covered in the eTTM. Therefore, to absorb the same energy content, the distribution is higher than for the other two models.

We distinguish between two types of multi-photon absorption: Successive or simultaneous absorption of two photons. 
Both processes allow depopulation of the energy region  $\left[\SI{-3.1}{\eV}, \SI{-1.55}{\eV}\right]$ and population of the energy region $\left[\SI{1.55}{\eV}, \SI{3.1}{\eV}\right]$.
Successively absorbing photons allows the excitation of electrons from $\left[\SI{-3.1}{\eV}, \SI{-1.55}{\eV}\right]$ to states below the Fermi edge which were initially filled.
This results in features of the DOS below the Fermi edge, which can be observed in the distributions of the Boltzmann model and AthEM. In our current implementation of the Boltzmann model, we additionally incorporate simultaneous absorption of two photons as described by Rethfeld \emph{et al.}~\cite{Rethfeld2002}, which leads to features of the DOS above the Fermi edge similar to the absorption of one photon with twice the energy.
It is less pronounced due to the low probability of this process. This is currently not captured in the AthEM model, which is why this feature is absent in AthEM. However, AthEM can in the future easily be extended to include such simultaneous multiphoton absorption effects.

%------------------------------------
\subsection{Total non-equilibrium distribution of gold after laser excitation}
\begin{figure}
    \centering
    \includegraphics{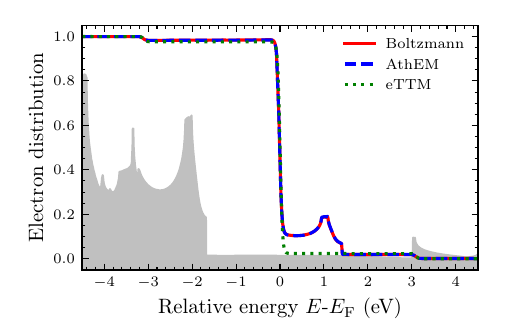}
    \caption{Total electron distributions as predicted by AthEM, eTTM~\cite{Uehlein2022}, and kinetic Boltzmann simulations with full collision integral for Au. Distributions are shown immediately after a laser pulse with a photon energy of \SI{3.1}{\eV}. Thermalization processes are not considered. The DOS for Au is sketched in gray and parameter set \cref{tab_laser}~B is used.}
    \label{fig:after_laser}
\end{figure}

\Cref{fig:after_laser} shows the total electron distribution for Gold calculated with Boltzmann collision integral, AthEM and eTTM at the end of a laser pulse with photon energy of \SI{3.1}{\eV}.

%------------------------------------
\subsection{Distributions of aluminum during thermalization}
\begin{figure}
    \centering
    \includegraphics{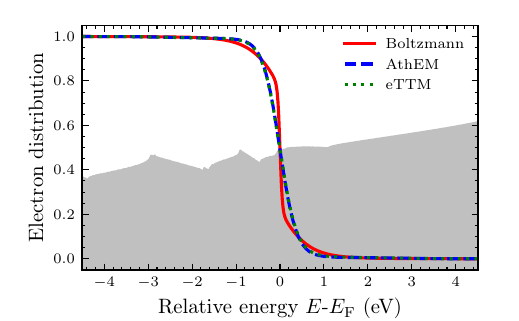}
    \caption{
    Non-equilibrium distributions of aluminum at the peak of the laser pulse.
    The results using Boltzmann collision integrals are compared to them resulting of the AthEM and eTTM.
    The shape of the DOS is sketched in gray and parameter set \cref{tab_laser}~A is used.
    All three graphs are similar.
    }
    \label{fig:al_dist}
\end{figure}
\begin{figure}
    \centering
    \includegraphics{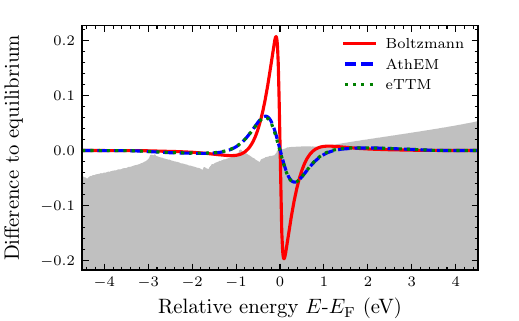}
    \caption{
    Difference of the non-equilibrium distributions of aluminum to the corresponding Fermi distribution at the peak of the laser pulse.
  The results using Boltzmann collision integrals are compared to them resulting of the AthEM and eTTM.
  The shape of the DOS is sketched in gray and parameter set \cref{tab_laser}~A is used.
  Due to electron-electron scattering, the distributions began to thermalize.
  All three graphs are similar.
  The curve from the Boltzmann model shows a stronger deviation from equilibrium than the other two curves.
    }
    \label{fig:al_diff}
\end{figure}

\Cref{fig:al_dist} shows the distribution of aluminum calculated with the three different models.
They are shown at the maximum of the laser pulse.
The features of the laser excitation are no longer visible, due to the thermalization, which has, in this fluence regime, already a strong influence during the laser pulse.
It is visible, that the Boltzmann model calculates a slower thermalization around the Fermi edge, leading to significant deviations from the Fermi distribution at this point in time.
Although the initial impression may be to the contrary, however, a detailed examination reveals that the other two distributions are also not Fermi distributed.

This is visible when plotting the difference between the distributions and a Fermi distribution with the same energy and particle content (\cref{fig:al_diff}). Comparing the three models, it is apparent, that the Boltzmann model shows the strongest deviation from equilibrium. However, the other two distributions also deviate from the equilibrium state.

In contrast to the results for gold in the main manuscript, no DOS features are visible for aluminum.

%%%%%%%%%%%%%%%%%%%%%%%%%%%%%%%%%%%%%

\section{Connecting dynamics of energy and particle density to temperature and chemical potential}

The differential equations for the thermal subsystems of our model describe the change in energy density~$u$ and particle density~$n$. 
To evaluate the energy resolved dynamics of all subsystems, we need the Fermi distribution of the thermal electron system and therefore their temperature~$T\el$ and chemical potential~$\mu$.
We use the dependence of the electron energy and particle density on the temperature and chemical potential to rewrite the change in energy and particle density as,
\begin{subequations}
\begin{align}
    \dv{u\el}{t} &= \pdv{u\el}{T\el} \dv{T\el}{t} + \pdv{u\el}{\mu} \dv{\mu}{t}\,,\label{eq:dudt}\\
    \dv{n\el}{t} &= \pdv{n\el}{T\el} \dv{T\el}{t} + \pdv{n\el}{\mu} \dv{\mu}{t}\,.\label{eq:dndt}
\end{align}
\end{subequations}
Furthermore, we introduce abbreviations for the partial derivatives with respect to temperature and chemical potential~\cite{Mueller2014PRB}, which follow from integration of the energy distribution $f\el$ 
\begin{subequations}
\begin{align}
	c_x &= \frac{\partial u\el}{\partial x} 
		= \int \frac{\partial f\el(E, T\el, \mu)}{\partial x}\,D(E)\,E\,\mathrm{d}E\,,\label{eq:c_x}\\
	p_x &= \frac{\partial n\el}{\partial x}
		= \int \frac{\partial f\el(E, T\el, \mu)}{\partial x}\,D(E)\,\mathrm{d}E\label{eq:p_x}
\end{align}
\end{subequations}
with $x \in \{T\el, \mu\}$ and using the electronic density of states~$D(E)$ of all considered electrons.
The integration in \cref{eq:c_x}  and \cref{eq:p_x} starts at the bottom of the considered bands,
and the partial derivatives of the Fermi distribution~$f\el(E, T\el, \mu)$ can be evaluated analytically.

With this, the equations~(\ref{eq:dudt}) and (\ref{eq:dndt}) are transformed to an equation for the temperature change,
\begin{align}
    \dv{T\el}{t} = \frac{1}{c_T p_\mu - p_T c_\mu} \left(p_\mu \dv{u\el}{t} - c_\mu \dv{n\el}{t} \right)\,.
\end{align}
The chemical potential is determined by a one-dimensional root-finding for the density of the electron subsystem applying the change in temperature and particle density.

For the phonons, we have 
\begin{align}
   \dv{u\ph}{t} = \pdv{u\ph}{T\ph}\,\dv{T\ph}{t} = c\ph \dv{T\ph}{t}\,,
\end{align}
where we assume a constant heat capacity $c\ph$ for the phonons.

%%%%%%%%%%%%%%%%%%%%%%%%%%%%%%%%%%%%%%%%%%%%%%%%%%%%%

	\bibliography{bibfile/Main.bib}